\documentclass[journal]{IEEEtran}
\usepackage{cite}

\ifCLASSINFOpdf
\else
   \usepackage[dvips]{graphicx}
\fi
\begin{document}
%
\title{A linear filter to reconstruct the ISW effect from CMB and LSS
observations}
%
%
%

\author{R. B.~Barreiro, P.~Vielva,
C.~Hern\'andez-Monteagudo and E.~Mart\'{\i}nez-Gonz\'alez%

\thanks{R.B. Barreiro, P. Vielva and E. Mart\'{\i}nez-Gonz\'alez are
at the Instituto de F\'\i sica de Cantabria (CSIC-Universidad de
Cantabria), Santander, Spain}

\thanks{Carlos~Hern\'andez-Monteagudo is at Max Planck Institut
f\"ur Astrophysik, Munich, Germany}

\thanks{Manuscript received --; revised --.}}

%
%

\markboth{Journal of \LaTeX\ Class Files --}%
{Shell \MakeLowercase{\textit{et al.}}: Bare Demo of IEEEtran.cls for Journals}
%



\maketitle

\begin{abstract}
The extraction of a signal from some observational data sets that
contain different contaminant emissions, often at a greater level than
the signal itself, is a common problem in Astrophysics and
Cosmology. The signal can be recovered, for instance, using a simple
Wiener filter. However, in certain cases, additional information may
also be available, such as a second observation which correlates to a
certain level with the sought signal. In order to improve the quality
of the reconstruction, it would be useful to include as well this
additional information. Under these circumstances, we have constructed
a linear filter, the linear covariance-based filter, that extracts the
signal from the data but takes also into account the correlation with
the second observation. To illustrate the performance of the method,
we present a simple application to reconstruct the so-called
Integrated Sachs-Wolfe effect from simulated observations of the
Cosmic Microwave Background and of catalogues of galaxies.

\end{abstract}


\IEEEpeerreviewmaketitle

\section{Introduction}
\IEEEPARstart{T}{HE} issues of signal reconstruction and component
separation have become of major importance in the field of
Astrophysics and Cosmology, covering a large range of applications.
In particular, Cosmic Microwave Background (CMB) observations require
the development of sophisticated image processing techniques, in most
cases on the sphere, to extract all the valuable information encoded
in this type of data. The CMB is a relic radiation that was emitted
shortly after the Big Bang, when the universe was $\sim$ 380000 years
old and has travelled since then -- almost unhindered -- towards us,
carrying a wealth of information about the origin and evolution of the
universe (see e.g. \cite{cha05,mar06} and references
therein). From the standard theory of inflation, the statistical
distribution of the CMB anisotropies is expected to follow an
isotropic and Gaussian random field on the sphere.
Different physical phenomena leave their imprint in the CMB
radiation that we observe today. In particular, a relevant
contribution is given by the so-called Integrated Sachs-Wolfe (ISW)
effect \cite{sac67}, which is due to the evolution in time of
the gravitational potential, in the linear regime, of the Large Scale
Structure (LSS) and its interaction with the CMB. This effect is of
great interest since it provides a direct indication of either the
presence of dark energy in the case of a flat universe or the
existence of spatial curvature \cite{pee03}.

Using only CMB observations, it is difficult to obtain a direct
detection of the ISW signal, since its contribution to the total CMB
signal is, in general, small, and is difficult to disentangle it from
other physical effects that also constitute the observed CMB
picture. However, given the origin of this effect, a spatial
correlation between the ISW and the LSS is expected.  Therefore, it
was suggested \cite{cri96} that the signal of the ISW effect could be
statistically detected by looking for cross-correlations between CMB
and LSS observations. Following these ideas, different groups have
obtained a statistical detection of the ISW effect using the CMB data
provided by the WMAP NASA satellite and different LSS surveys at
levels between $2.5\sigma$ and $4.5\sigma$
\cite{fos03,bou04,fos04,afs04,nol04,vie06,pie06,mce07,mce08,rac08,gia08}.

Nonetheless, it would be desirable to obtain not only a statistical
detection of the ISW signal but also the actual map of this effect. A
possible procedure would be to apply a reconstruction image technique
(e.g. Wiener filter \cite{wie49}, maximum-entropy method \cite{ski89})
to the CMB map in order to extract the ISW effect from the rest of
contributions of the primordial radiation. However, given the weak
level of the signal such a simple procedure is not expected to provide
good results. Also, techniques that make use of multifrequency
observations (see \cite{del08} and references therein) are not useful
since the signal has the same frequency dependence as the other
dominant CMB contributions. Thus, instead, we propose to recover the
ISW map applying a filter, that we have called linear covariance-based
(LCB) filter, to the CMB data that takes also into account the
information coming from the cross-correlation between the CMB and the
LSS surveys.

The outline of the paper is as follows. Section~\ref{sec:method}
presents our method to reconstruct a signal embedded in a noisy
background and correlated with a second
observation. Section~\ref{sec:isw} shows the performance of the
proposed LCB filter in a simple application, the recovery of the ISW
signal from simulated data sets. The robustness of the
methodology against certain non-idealities is studied in
section~\ref{sec:robustness}. Finally, we present our conclusions in
section~\ref{sec:conclusions}.

\section{The method}
\label{sec:method}

A random signal $a$ measured on the sphere, such as the CMB,
is usually expanded in terms of the spherical harmonics $Y_{\ell m}$
\begin{equation}
a(\theta, \phi)=\sum_{l=0}^{\infty} \sum_{m=-\ell}^{\ell} a_{\ell m}
Y_{\ell m}(\theta, \phi)
\end{equation}
where $\theta \in [0, \pi]$ and $\phi \in [0, 2\pi]$ stand for the
colatitude and longitude in spherical coordinates, respectively, and
$a_{\ell m}$ are the coefficients of the expansion. For a
statistically isotropic signal, the variance of the coefficients is
independent of $m$:
\begin{equation}
\left< a_{\ell m} a^*_{\ell' m'} \right> = C_{\ell} \delta_{\ell
\ell'} \delta_{m m'}
\end{equation}
where the averages are to be taken over statistical ensembles and
the $C_\ell$'s constitute the angular power spectrum. For the case
of an isotropic Gaussian random field on the sphere, the power
spectrum completely characterises the signal.

\subsection{Wiener filter (WF)}

Let us assume that we have an observation $d$ that contains a certain
signal of interest $s$ plus some generic noise $n$, both of which are
assumed to be Gaussian and isotropic. One possible way to obtain an
estimation $\hat{s}$ of the signal is to apply the classical WF to the
data, which, at each $(\ell, m)$ mode, is given by:
\begin{equation}
\hat{s}_{\ell m}=\frac{C_\ell^s}{C_\ell^s+C_\ell^n}d_{\ell m}
\end{equation}
where $C_\ell^s$ and $C_\ell^n$ correspond to the power spectrum of
signal and noise respectively. It can be shown that WF
is the linear filter that minimises the variance of the reconstruction
error.

It is well known that the power spectrum of the WF reconstruction
is biased towards values lower than the true signal, with the
bias depending on the signal-to-noise ratio of the data. In
particular, from the previous equation, it is straightforward to show
that the expected value of the power spectrum for the reconstructed
signal is given by
\begin{equation}
\label{eq:cl_wf}
\left <C_\ell^{\hat{s}} \right> = \frac{(C_\ell^s)^2}{C_\ell^s+C_\ell^n}
\end{equation}
or, alternatively, we can write the expected bias of the signal as 
\begin{equation}
\left < b \right> =
C_\ell^s - \left <C_\ell^{\hat{s}} \right> 
=  \frac{C_{\ell}^s C_{\ell}^n}{C_{\ell}^s + C_{\ell}^n }
\label{eq:bias_wf}
\end{equation}
Note that in the limit of infinite signal-to-noise ratio, the WF goes
to unity and the bias becomes negligible.

\subsection{Linear covariance-based (LCB) filter}

Let us assume that we have an additional second observation $g$ which
is also isotropic and Gaussian and that is correlated with the signal
$s$. It would be desirable to include the information of $g$ in order
to reconstruct $s$. The signals are characterised by their angular
power spectra and also by their cross power spectrum:
\begin{equation}
C_{\ell}^{gs} = \left< g_{\ell m} s^*_{\ell' m'} \right>
\end{equation}
which we assume to be (statistically) known\footnote{In the particular
case of the ISW recovery, this corresponds to the expected
cross-correlation between the ISW signal and the projected galaxy
density. This quantity is given in terms of cosmological parameters
that define the expansion and the energy content of the Universe, and
the astrophysical laws that guide the LSS evolution.}. The covariance
matrix $C(\ell)$ of $g$ and $s$ at each multipole $\ell$ is
constituted by the expected power spectra of the signals (diagonal
elements) and their cross-power spectrum (off-diagonal elements). It
is useful to calculate the Cholesky decomposition of the covariance
matrix, which satisfies $C(\ell)=L(\ell)L^T(\ell)$, where $L$ is a
lower triangular matrix.  It can be shown that $g_{\ell m}$ and
$s_{\ell m}$ can be written as
\begin{equation}
\label{eq:hidden}
\left(
\begin{array}{c}
g_{\ell m} \\
s_{\ell m} \\
\end{array}
\right)
=
\left(
\begin{array}{cc}
L_{11}(\ell) & 0 \\
L_{12}(\ell) & L_{22} (\ell)\\
\end{array}
\right)
\left(
\begin{array}{c}
h_{\ell m} \\
j_{\ell m} \\
\end{array}
\right)
\end{equation}
where $h$ and $j$ are uncorrelated Gaussian variables of unit
variance, sometimes referred to as hidden variables. Note that the
elements of $L(\ell)$ relate to the elements of the covariance matrix
$C(\ell)$ as $L_{11}=\sqrt{C_\ell^g},
L_{12}=C_\ell^{gs}/\sqrt{C_\ell^g}$ and
$L_{22}=\sqrt{\left|C(\ell)\right|/C_\ell^g}$, where
$\left|C(\ell)\right|$ is the determinant of the covariance matrix at
each $\ell$ mode.

From the previous equation, we can trivially write $s$ as a function
of $g$ and $j$ as
\begin{equation}
\label{eq:signal}
s_{\ell m}=\frac{L_{12}(\ell)}{L_{11}(\ell)}g_{\ell m} + L_{22}(\ell)j_{\ell m}
\end{equation}
Let us recall the observational data $d$, which can now be written as
\begin{equation}
d_{\ell m}=s_{\ell m}+n_{\ell m}=\frac{L_{12}(\ell)}{L_{11}(\ell)}g_{\ell m} + L_{22}(\ell)j_{\ell m}+n_{\ell m}
\end{equation}
In order to recover $s$, we can now proceed in the following way. The
first term of equation (\ref{eq:signal}) can be easily calculated from
$g$ and the covariance matrix of the signals. The second term can be
estimated by applying WF to the modified data $\bar{d}$
given by
\begin{equation}
\bar{d}_{\ell m}=d_{\ell m}-\frac{L_{12}(\ell)}{L_{11}(\ell)}g_{\ell
m}= L_{22}(\ell)j_{\ell m}+n_{\ell m}
\end{equation}
Therefore, the final estimation $\hat{s}$, at each multipole, is obtained as
\begin{equation}
\label{eq:rec}
\hat{s}_{\ell m}=\frac{L_{12}(\ell)}{L_{11}(\ell)}g_{\ell m} + \frac{L_{22}^2(\ell)}{L_{22}^2(\ell)+C_{\ell}^n}
\bar{d}_{\ell m}
\end{equation}
This simple procedure allows one to take into account the second
correlated observation $g$ in the reconstruction of $s$. This improves
the quality of the reconstructed signal that also has the right
correlation with $g$.

It can be easily shown that the expected value of the power
spectrum for the reconstructed signal is given by
\begin{equation}
\label{eq:cl_lcb}
\left< C_\ell^{\hat{s}} \right>= \frac{(C_\ell^{gs})^2\left(\left|C(\ell)\right|+C_\ell^g
C_\ell^n\right)+\left|
C(\ell)\right|^2}{C_\ell^g\left(\left|C(\ell)\right|+C_\ell^g C_\ell^n\right)}
\end{equation}
As we are using WF to recover a part of the signal, the power spectrum
of the reconstruction is again biased.  However, it can be shown that
this bias is always lower or equal than the one of the WF
reconstruction. To prove this statement, we first note 
that the cross power spectrum must satisfy $(C_\ell^{gs})^2= \alpha C_\ell^{g}
C_\ell^{s}$, with $0 \le \alpha \le 1$ . Taking
this property into account and making use of
equation~(\ref{eq:cl_lcb}), we find that the bias introduced by the
LCB filter is given by
\begin{equation}
\left< b \right> = \frac{C_{\ell}^s C_{\ell}^n}{C_{\ell}^s + \frac{C_{\ell}^n}{1-\alpha} }
\label{eq:bias}
\end{equation}
As one would expect, the bias is a function of ${C_{\ell}^s}$,
${C_{\ell}^n}$ and $\alpha$, but does not depend on ${C_{\ell}^g}$. It
is interesting to point out that for $\alpha=0$, the LCB filter
defaults to WF, since in this case $g$ does not contain information
about $s$. For $\alpha=1$, the bias goes to zero, and the
reconstructed signal is simply proportional to the observation
$g$. Also, by comparing equations (\ref{eq:bias_wf}) and
(\ref{eq:bias}), and since $0 \le \alpha \le 1 $, it becomes apparent
that the bias of the LCB filter reconstruction is always lower or
equal than the one of WF.

Finally, we would like to remark that the described procedure can be
easily generalised to include additional correlated signals, providing
the expected value of the full covariance matrix is known.

\section{Reconstructing the ISW map}
\label{sec:isw}

In order to test the performance of our method, we have considered the
problem of reconstructing the ISW map from CMB and LSS
observations. In our case, the signal $s$ corresponds to the
temperature fluctuations caused by the ISW effect, the observed data
$d$ is a CMB observation, the noise $n$ is all the signal present in
the CMB map that is not ISW and $g$ is a catalogue of galaxies (i.e. a
map of the galaxy density projected along the line of sight) which is
correlated with the ISW pattern through the gravitational potential
which affects to both. The reconstruction of the ISW map is a very
challenging problem since the signal is very weak, in particular the
signal-to-noise ratio for the considered example (maps with $\sim$
2-degree resolution) is around 0.4. Although real CMB and LSS data are
affected by non-idealities, like incomplete sky coverage and residual
signals coming from astrophysical contaminants, in this section we
will consider full-sky observations free of residual contaminants. The
reason is that this simpler scenario allows one to understand better
the performance of the LCB filter. However, in section
\ref{sec:robustness} we will study the robustness of the method in
more realistic situations.

We have generated correlated simulations of CMB and catalogues of
galaxies making use of equation (\ref{eq:hidden}) (see also
\cite{bou98}). More specifically, for a given model and at each
multipole, first we calculate the covariance matrix $C(\ell)$ (using a
modified version of CMBFAST \cite{sel96}) and its corresponding
Cholesky decomposition $L(\ell)$. Then, we generate two uncorrelated
Gaussian variables, $h_{\ell m}$ and $j_{\ell m}$, of zero mean and
unit dispersion. These quantities are then combined following equation
(\ref{eq:hidden}) to obtain the coherent ISW and galaxy catalogue
simulations. Finally, an independent Gaussian CMB simulation is
generated (without including the ISW contribution) and added to the
ISW map to obtain the simulated data $d$.  Since the ISW effect is
important at large angular scales, our simulated maps have a
resolution of $\sim$ 2 degrees, corresponding to the
HEALPix\footnote{HEALPix is a hierarchical equal area isolatitude
pixelization of the sphere. For more information see
http://healpix.jpl.nasa.gov/} parameter $n_{side}=32$. For the CMB
power spectra the flat $\Lambda$-CDM model that best fits the 3-year
WMAP data \cite{spe07} has been used.  According to this model, the
Universe has undergone a recent epoch of accelerated expansion, and
the energy causing this acceleration amounts, at the present time,
around 75 per cent of the total energy budget of the Universe. We have
neither included a beam nor instrumental noise, since, for the current
and future CMB observations, their effect is negligible at the
considered angular scales. For the galaxy catalogue, we have generated
several idealised surveys centred at different redshifts $z$ (an
astrophysical parameter related to the distance of the galaxies from
the observer).  It is known that galaxies are biased tracers of the
underlying matter density field present in the Universe, in the sense
that they tend to populate most of the times the most overdense
regions of the Universe. This bias is dependent of the galaxy type,
and a priori may depend as well on the age of the Universe (also
related to the redshift parameter $z$).  For simplicity we have
considered a constant bias and a population of galaxies with no
intrinsic evolution with redshift. It can be shown from
equation~(\ref{eq:rec}) that the reconstruction of the signal is
independent of a constant bias, since it acts as a global
normalisation of the galaxy catalogue. The cross-power spectra between
the CMB and the different catalogues of galaxies have been obtained
following linear theory within the frame of the concordance
$\Lambda$CDM cosmological model. This assumes that perturbations in
the density field and gravitational potential are {\em small}, which
is appropriate for the large scales relevant to the ISW effect (we
know that the universe is isotropic and homogeneous in the very large
scales). This computation traces the freeze-out of the growth of the
gravitational potential wells due to the onset of an accelerated
expansion in the Universe, and couples it to the gravitational energy
that CMB photons gain or lose when crossing those potential wells.

\begin{figure}
\centering
\includegraphics[width=3.2in]{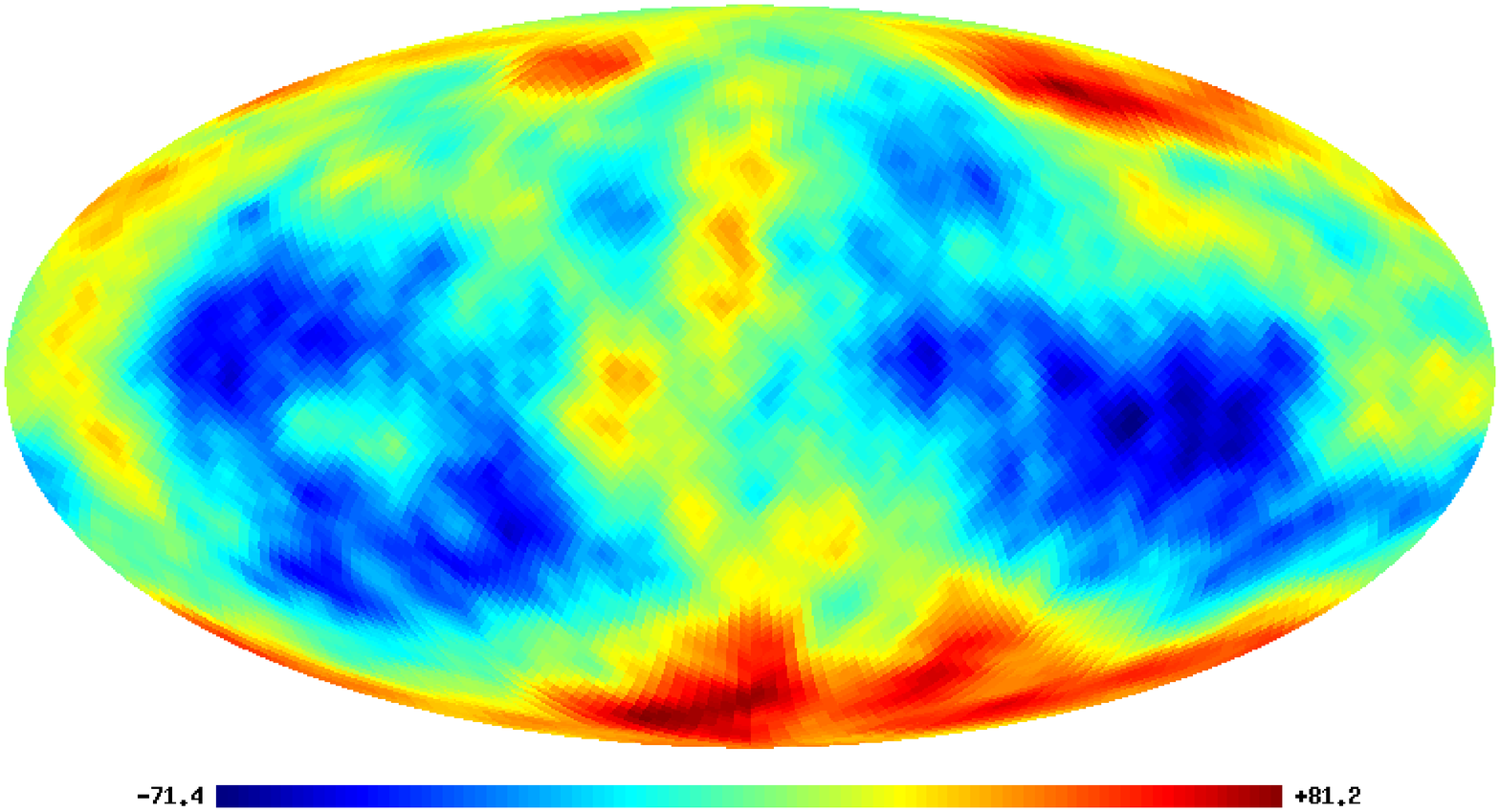}\\
\includegraphics[width=3.2in]{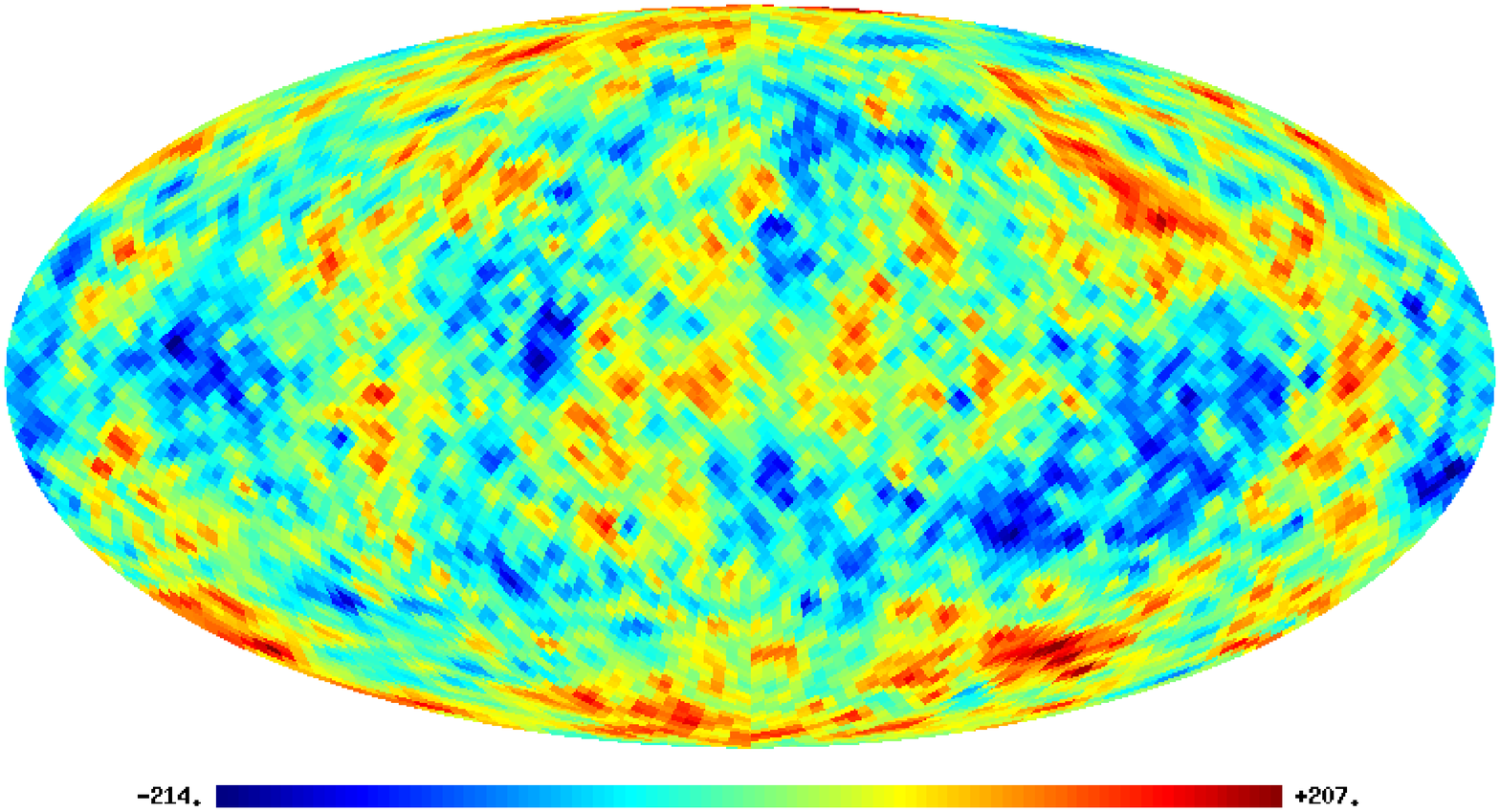}\\
\includegraphics[width=3.2in]{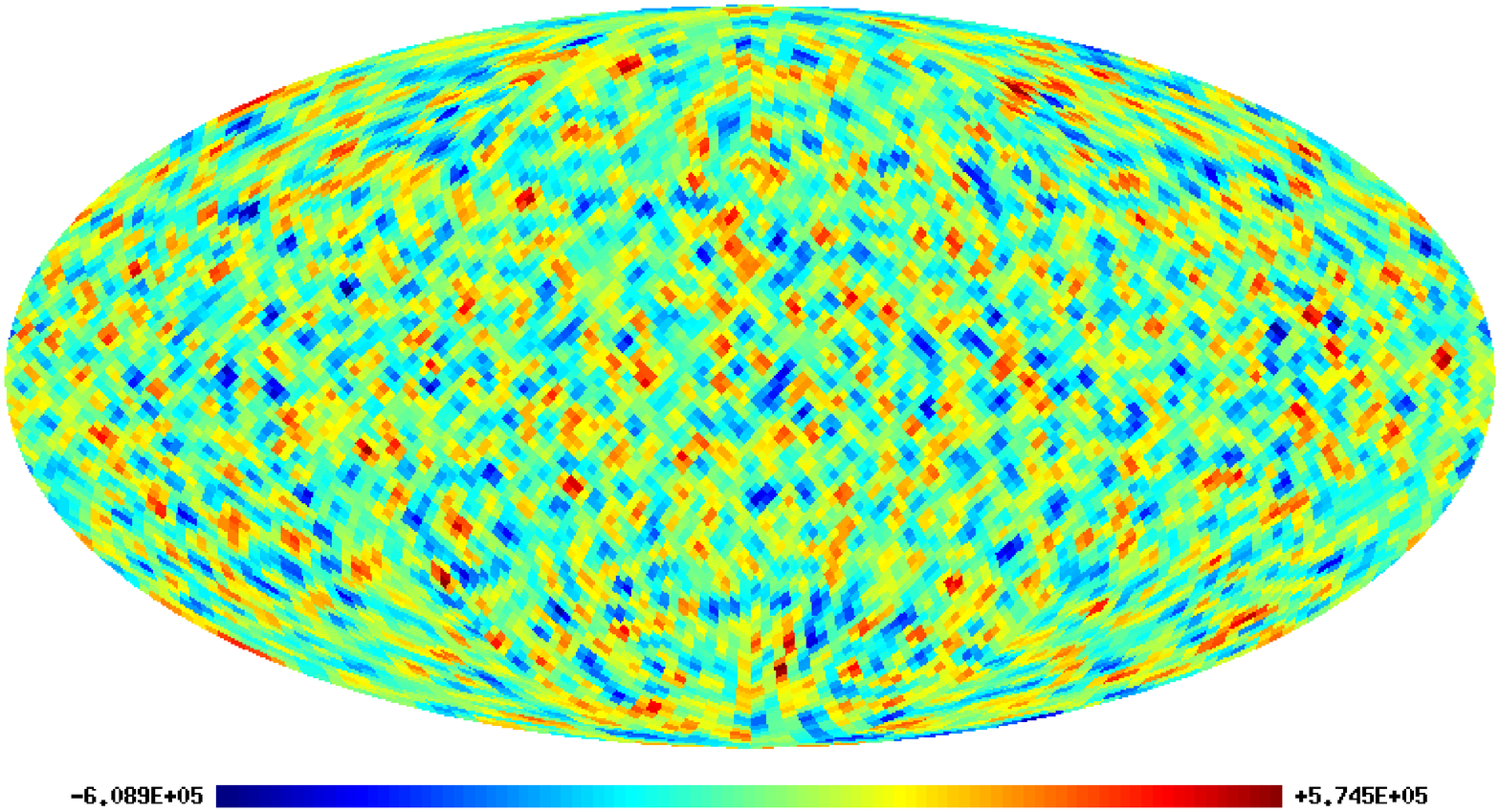}
\caption{The figure shows coherent simulations of an ISW map (top
panel), a CMB map including the ISW effect (middle panel) and a
catalogue of galaxies (bottom panel). The catalogue of galaxies has
arbitrary units (the reconstruction is independent of the
normalisation) and is centred at a comoving radial distance $r$
corresponding to a redshift $z = 0.8$. The width of the radial shell
containing the galaxies is approximately $0.1r$. The units of the CMB
and ISW maps are microkelvin.}
\label{fig:maps}
\end{figure}

\begin{figure}
\centering
\includegraphics[width=3.2in]{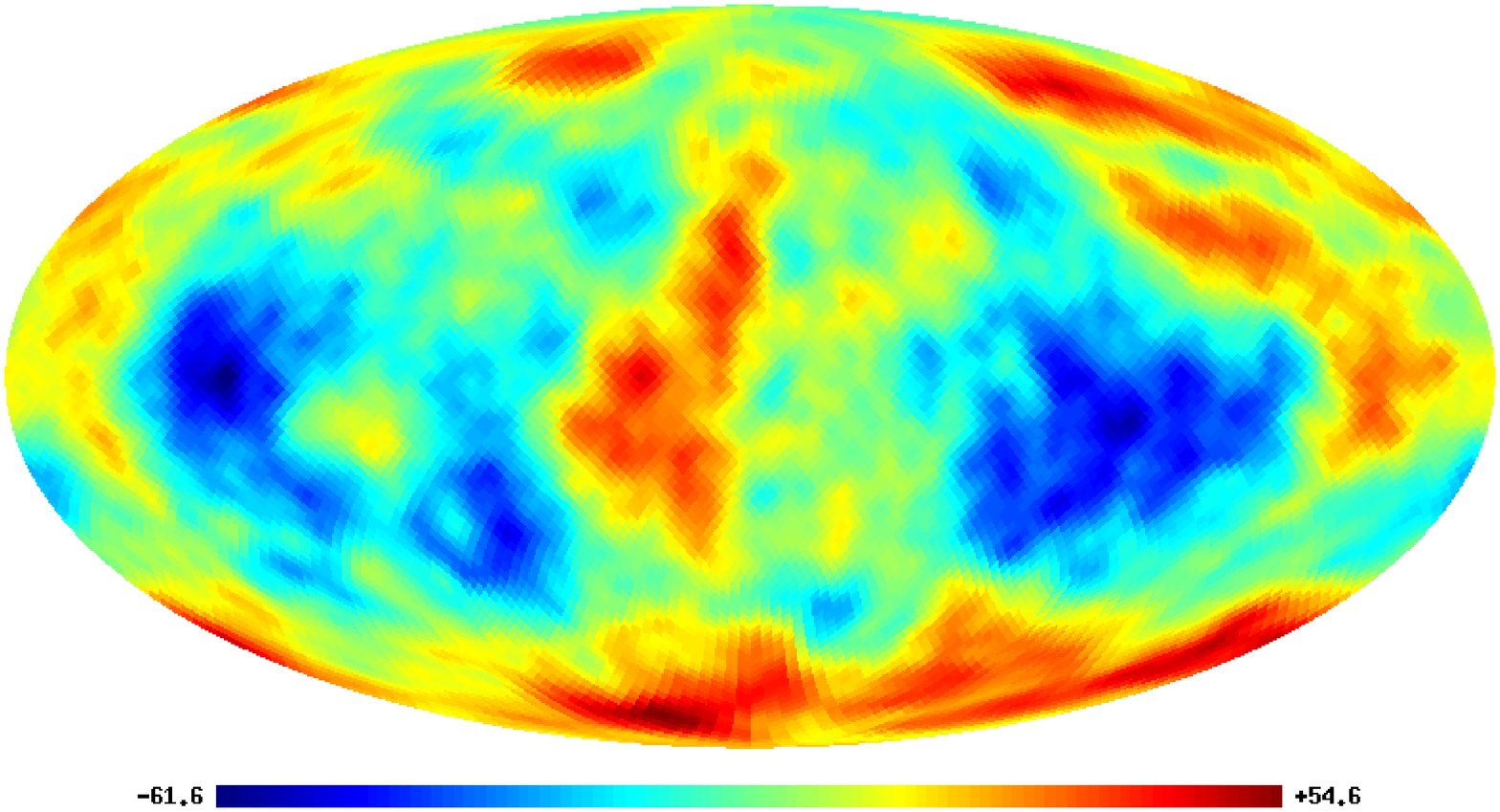} \\
\includegraphics[width=3.2in]{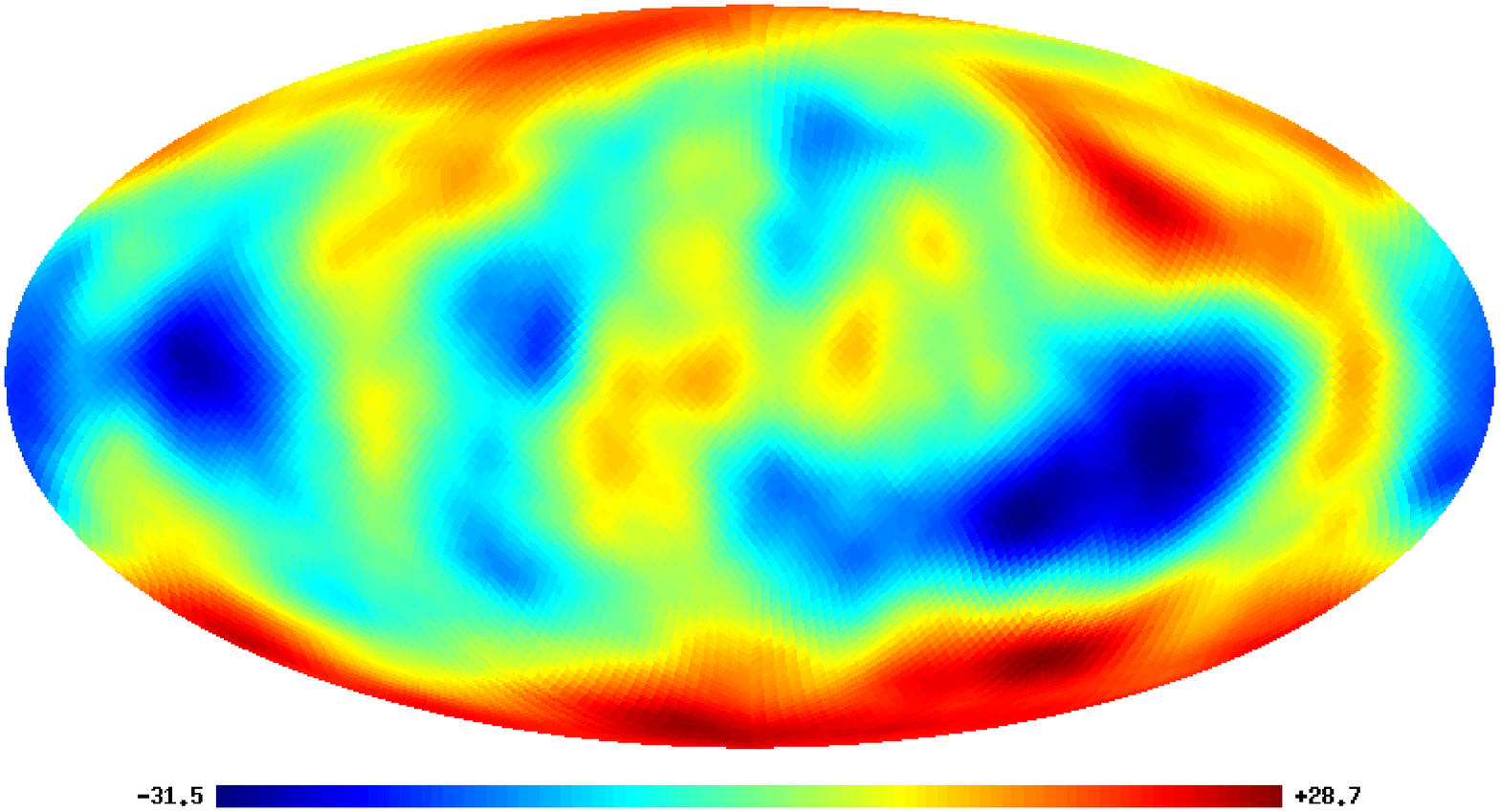}
\caption{Reconstruction of the ISW signal from the simulated data of
  Fig.~\ref{fig:maps} using the LCB filter (top) and the WF
  (bottom). The figures should be compared to the input ISW
  (top panel of Fig.~\ref{fig:maps}). The cross-correlation
coefficient $\rho$ is equal to 0.88 and 0.63 for the LCB and WF
reconstructions, respectively.}
\label{fig:rec}
\end{figure}

As a measurement of the quality of the recovered signal, we will use
the correlation coefficient $\rho$ between the input and reconstructed
maps which is defined as
\begin{equation}
\rho=\frac{\left<s \hat{s}\right> - \left< s \right>\left< \hat{s}
\right>}{\sigma_s \sigma_{\hat{s}}} ,
\end{equation}
where $\sigma_s$ and $\sigma_{\hat{s}}$ are the dispersion of the
input and reconstructed map respectively. Values of $\rho$ closer to
one provide better reconstructions. As an illustration,
Fig.~\ref{fig:maps} shows a simulation of the ISW, the CMB (including
the ISW) and a catalogue of galaxies centred at a comoving radial
distance $r$ corresponding to a redshift $z = 0.8$. The width of the
radial shell containing the galaxies is roughly $0.1r$.
Fig.~\ref{fig:rec} shows the reconstructed ISW map using the LCB
filter (top panel) for the set of simulations given in
Fig.~\ref{fig:maps}. For comparison, the WF reconstruction is also
shown (bottom panel). It is apparent that the map recovered with the
LCB filter follows better the general structure of the input ISW map
compared to that obtained with WF. This can also be seen from the
cross-correlation coefficient between the input and reconstructed map
which is 0.88 and 0.63 for the LCB and WF, respectively.

\begin{figure*}
\centering
\includegraphics[width=3.2in]{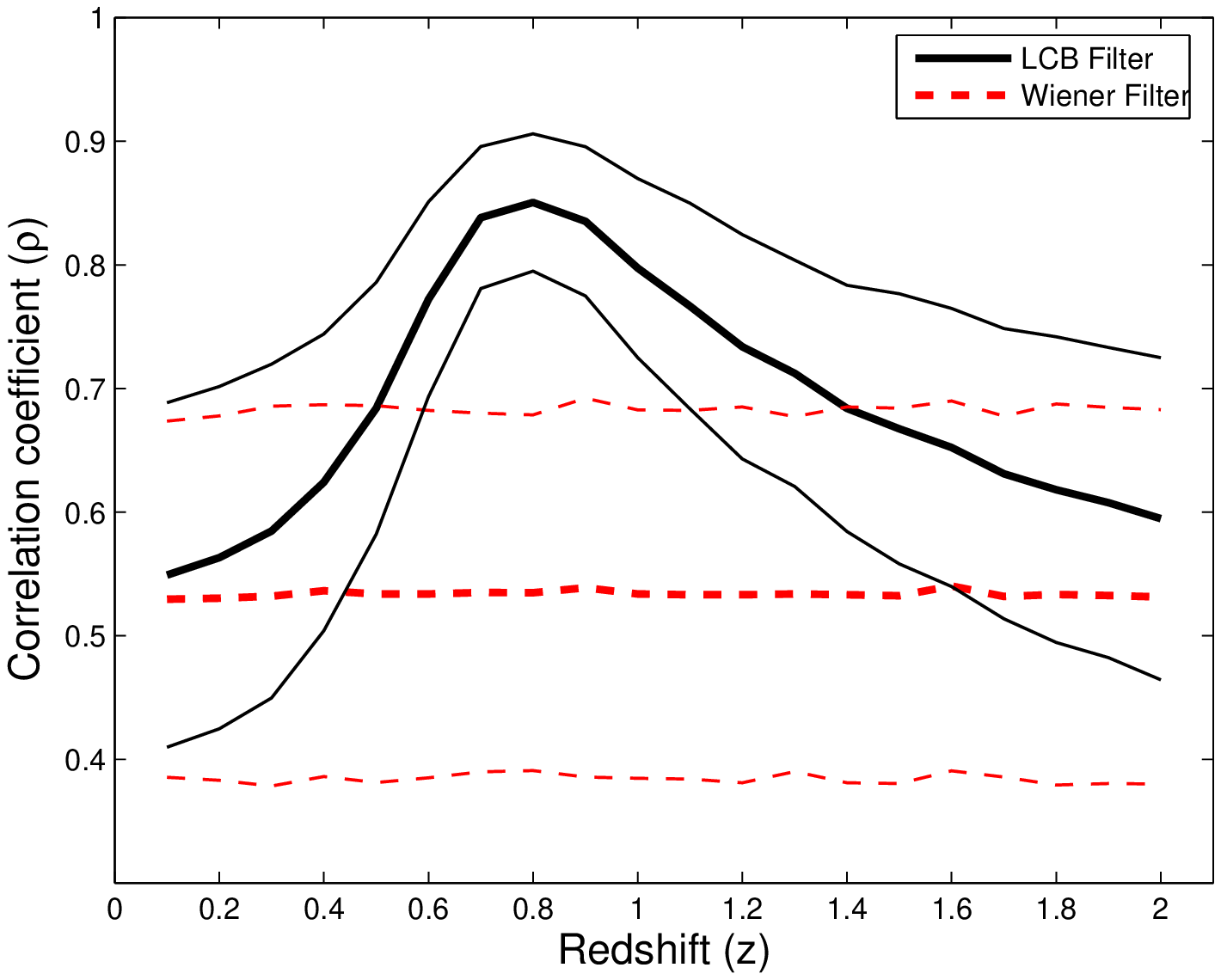}
\includegraphics[width=3.2in]{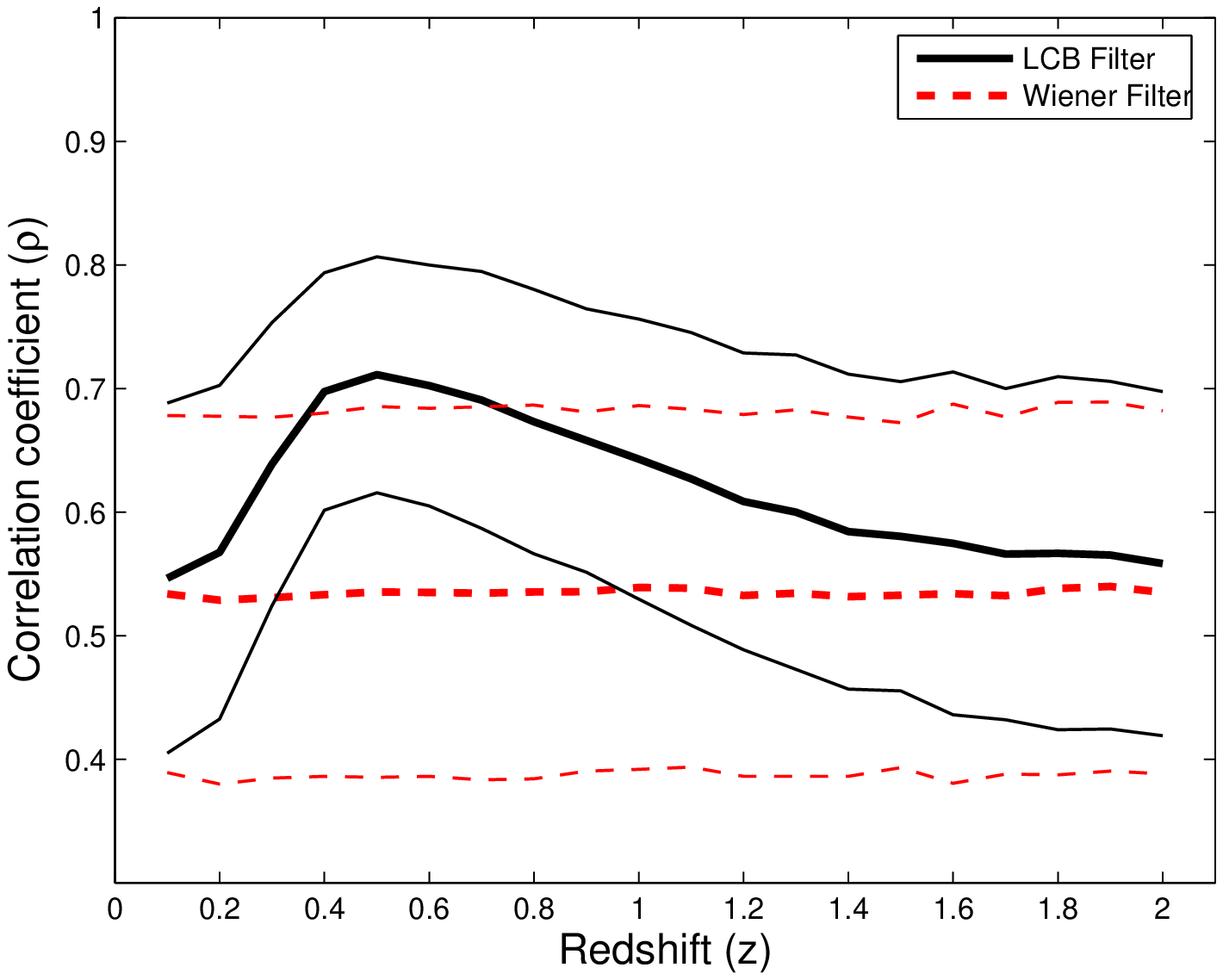}
\caption{Mean correlation coefficient for the ISW reconstruction using
the LCB filter obtained from 1000 simulations (black solid thick
line) and the corresponding 1$\sigma$ region. Two cases are
considered: a wider bin in $z$ (left panel) and a narrower bin (right
panel). For comparison the mean correlation coefficient and the
corresponding 1$\sigma$ region for the WF reconstruction
are also shown (red dashed lines).}
\label{fig:errors}
\end{figure*}

For a given CMB observation, the quality of the recovered ISW map
obtained with the LCB filter could be significantly different
depending on the catalogue of galaxies chosen for the reconstruction,
since the cross-correlation between the ISW and the LSS is redshift
dependent (as well as depends on the particular type of galaxies, the
completeness of the survey, etc.). To provide an insight of the
importance of this effect on the quality of the reconstruction,
Fig.~\ref{fig:errors} shows the correlation coefficient (averaged over
1000 simulations) for the LCB filter using a CMB map and different
LSS surveys, centred at redshift $z$ (black solid thick line). We have
also considered two widths in the range of redshifts observed by the
survey: a wider bin centred on $z$ (or on distance $r$) with a width
of $0.1r$ (left panel) and a narrower bin of width $0.05r$ (right
panel). The thinner solid curves provide the 1$\sigma$ region around
the mean $\rho$ obtained from the simulations. For comparison, the
results for the reconstruction from WF (red dashed thick
line) and the corresponding 1$\sigma$ region (red dashed thin lines)
are also shown. Note that the WF reconstruction is independent of
the considered LSS survey, since it does not use this information (the
small deviations from a constant line are due to the finite number of
simulations).
From Fig.~\ref{fig:errors} it is apparent that the LCB filter
outperforms the WF in all the considered cases.  One can
also notice that the surveys with a wider bin in $z$ work better than
those with a narrower bin. In particular, for the considered cases, we
find that the best reconstruction is obtained for a survey centred in
$z=0.8$ for the case of the wider bin. This indicates that having a
larger number of observed galaxies increases the correlation between
the ISW and the considered survey and that, therefore, the
reconstruction of the ISW signal is improved. However, we should
remark that the bias between the galaxy and the underlying matter
distribution is more unlikely to remain constant (as we have assumed)
for wide bins. For this reason, a more detailed study of the bias
within the redshift/distance bin would be required.

It is also interesting to point out that the quality of the
reconstruction is higher in the hottest and coldest spots of the ISW
map, since the signal is comparatively larger in those regions. For
instance, for the best reconstruction case, the average correlation
coefficient increases from $\rho = 0.85$, when all the pixels are
considered, to $\rho = 0.94$, when only the pixels outside the
1$\sigma$ region are considered.

\begin{figure}
\centering
\includegraphics[width=3.2in]{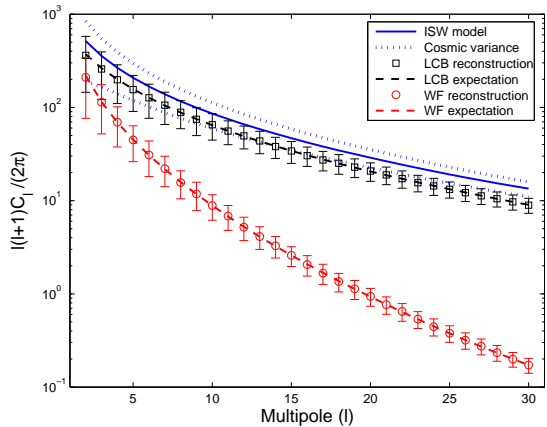}
\caption{Mean power spectrum of the ISW reconstruction using the LCB
filter obtained from 1000 simulations (black squares) compared to the
input model (blue solid line). The blue dotted lines indicate the
1$\sigma$ region allowed by the cosmic variance and the error bars
give the dispersion of the reconstructed power spectrum obtained from
simulations. The theoretical expectation for the power spectrum of
the reconstruction is also plotted (black dashed line). For
comparison, the same quantities are given for the Wiener filter case
(red circles and red dashed line). The reconstructions correspond to
the wider bin case, centred at $z=0.8$.  }
\label{fig:cls}
\end{figure}

We have also studied the power spectrum of the recovered ISW map,
which provides an alternative measurement of the quality of the
reconstruction. As mentioned in section~\ref{sec:method}, we expect the
power spectrum of our recovered ISW map to be biased at a certain
level. Fig.~\ref{fig:cls} shows the power spectrum of the
reconstructed ISW using the LCB filter averaged over 1000 simulations
(black squares) for the case of a catalogue of galaxies covering a
wide bin centred at redshift $z=0.8$, compared to the input ISW model
(blue solid line). The blue dotted lines correspond to the theoretical
cosmic variance while the error bars give the dispersion obtained from
simulations. For comparison, the expected reconstructed power spectrum
given by equation (\ref{eq:cl_lcb}) is also plotted (black dashed
line), showing an excellent agreement with the results obtained from
simulations. 
Although, as expected, the reconstructed power spectrum is biased,
taking into account the weakness of the ISW effect, a significant part
of the power of the signal is present in the reconstructed map. In
particular, the average reconstructed power spectrum lies within the
allowed $1\sigma$ cosmic variance region up to $\ell \sim 15$. We have
also studied, for the same case, the average power spectrum and
dispersion of the reconstructed ISW map using WF (red circles and
error bars in Fig.~\ref{fig:cls}), finding a much larger bias, in
agreement with equation (\ref{eq:cl_wf}) (red dashed line). This shows
again that combining information from the CMB and LSS, as in the LCB
filter, is very useful to recover the ISW map.

\section{Robustness of the method}
\label{sec:robustness}

In the previous section, we have shown the performance of the LCB
filter to recover the ISW effect from CMB and LSS observations in
ideal conditions. However, when dealing with real data, they will be
affected by different non-idealities, that should be taken into
account in the analysis. Regarding CMB data, one of the most important
issues is the contamination due to different astrophysical emissions,
which is particularly strong in the Galactic plane. This region is
usually discarded from the data, by applying a Galactic mask (i.e.
replacing with zeroes the most contaminated regions). Regarding
catalogues of galaxies, they do not usually cover the whole sky and,
therefore, it is necessary to deal with incomplete sky maps. 

\begin{figure}
\centering
\includegraphics[width=3.2in]{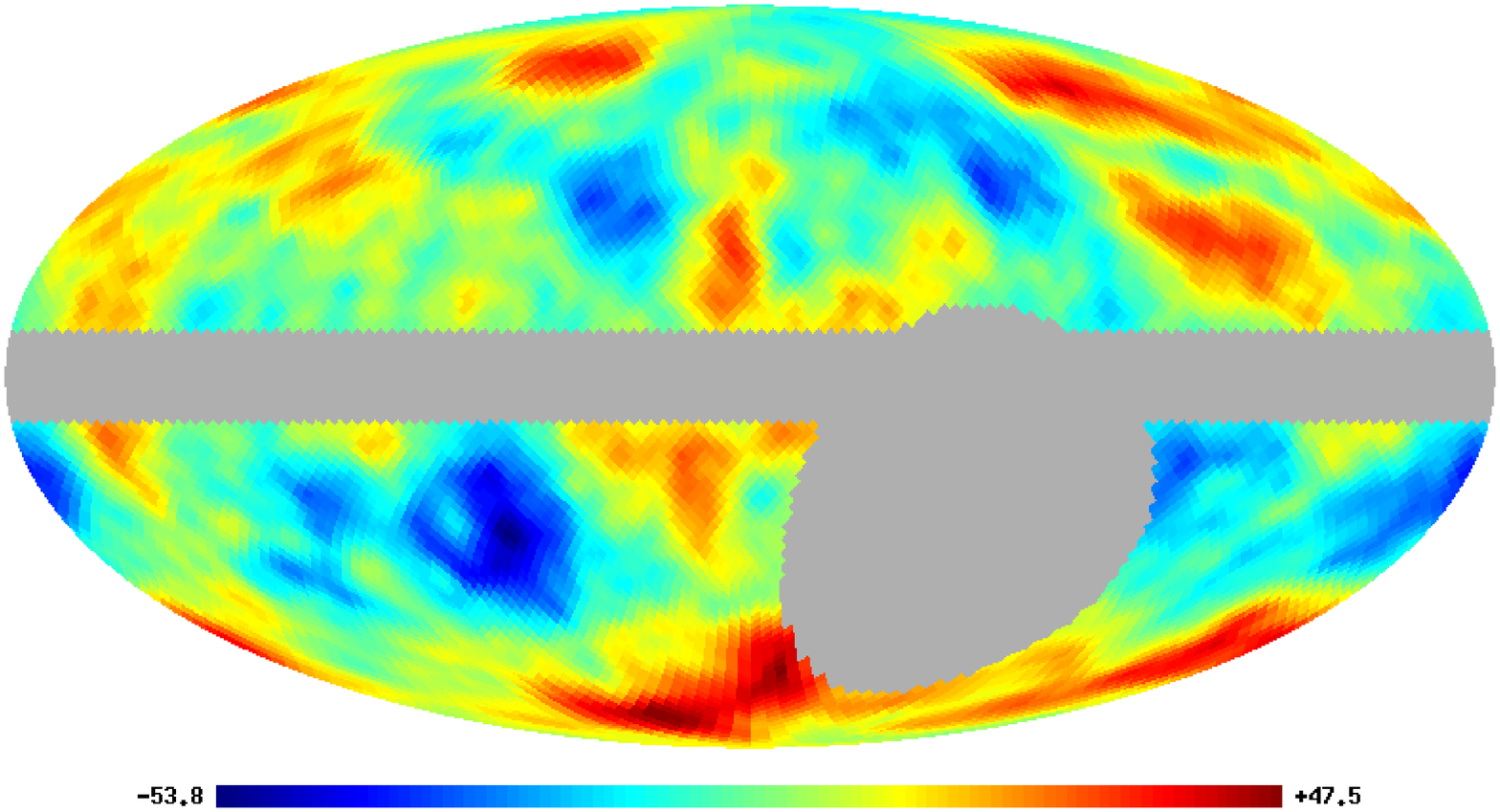} \\
\includegraphics[width=3.2in]{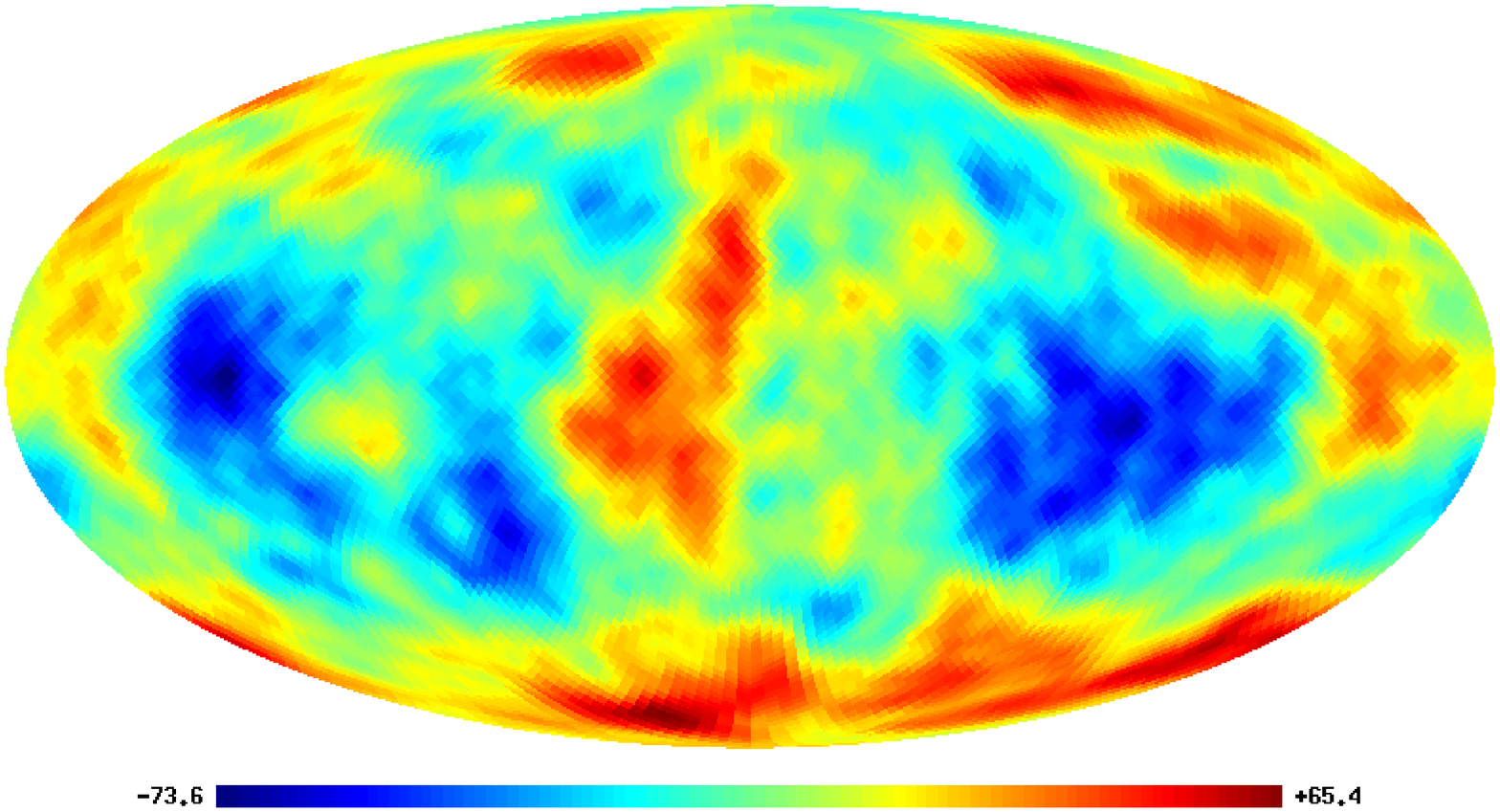}
\caption{Reconstruction of the ISW signal for the case of an
incomplete sky (top) and when assuming an erroneous cosmological model
(bottom). The figures should be compared to the input ISW (top panel
of Fig.~\ref{fig:maps}) and also to the reconstruction in the ideal
case (top panel of Fig.~\ref{fig:rec}). The grey region of the top
panel indicates the masked region, excluded from the analysis. The
cross-correlation coefficient $\rho$ is equal to 0.81 (top
reconstruction) and 0.87 (bottom reconstruction).}
\label{fig:rec_robust}
\end{figure}

In order to test the robustness of the LCB filter on an incomplete
sky, we have repeated the analysis from the previous section after
applying a simple mask to the data that covers around 25 per cent of
the sky (see top panel of Fig.~\ref{fig:rec_robust}). The mask has
been constructed as the addition of a band centred in the equator
with latitude $|b| \le 8^\circ$ plus a circle centred in the southern
ecliptic circle with a radius of 40$^\circ$. The equatorial band
mimics a simple Galactic mask from a CMB observation, discarding $\sim
15$ per cent of the sky, which is similar, for instance, to the area
covered by the WMAP Kp2 mask \cite{ben03}. The circle mimics a
region not observed by an incomplete galaxy survey, as in the case of
the NVSS catalogue \cite{con98}.

\begin{figure*}
\centering
\includegraphics[width=3.2in]{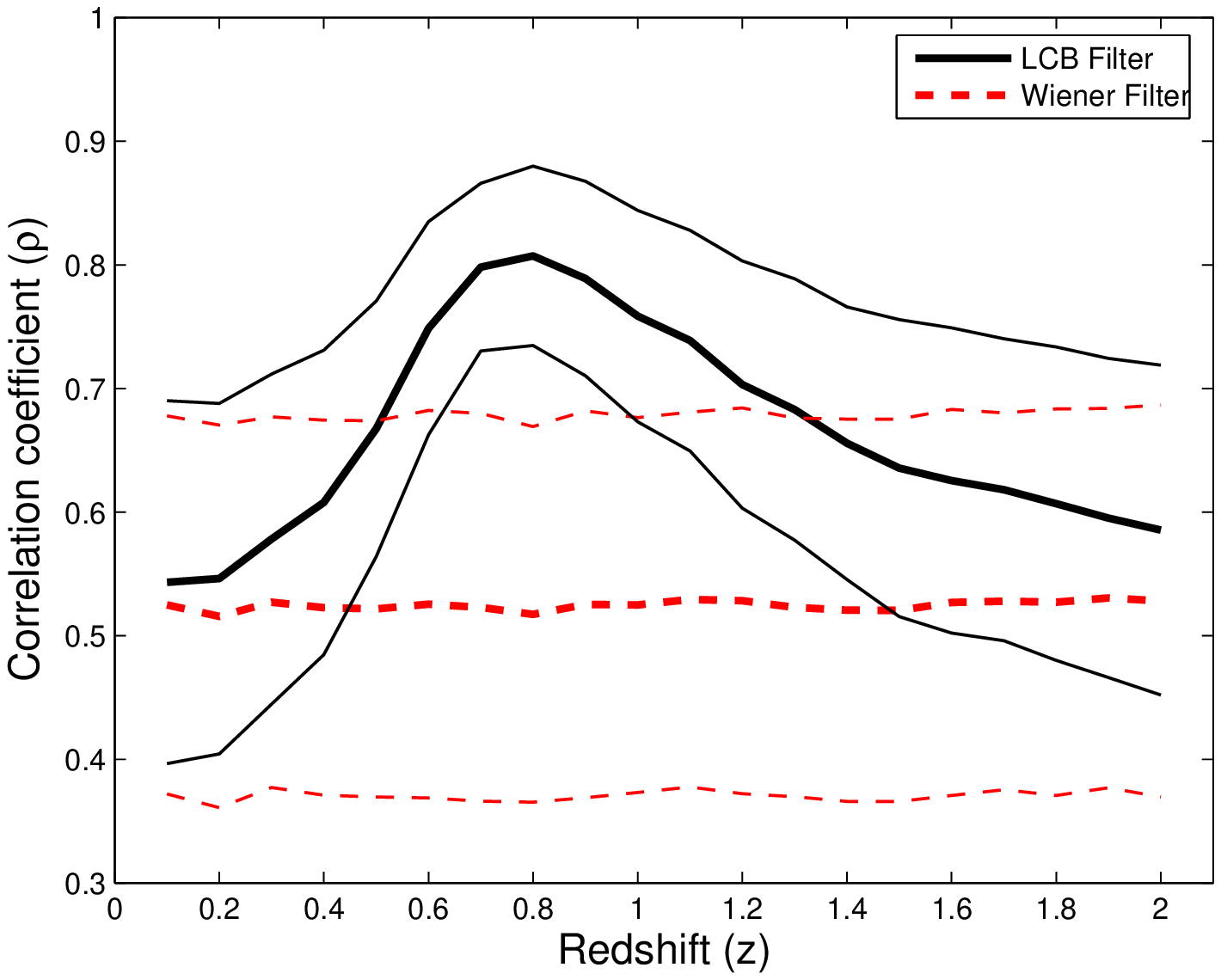}
\includegraphics[width=3.2in]{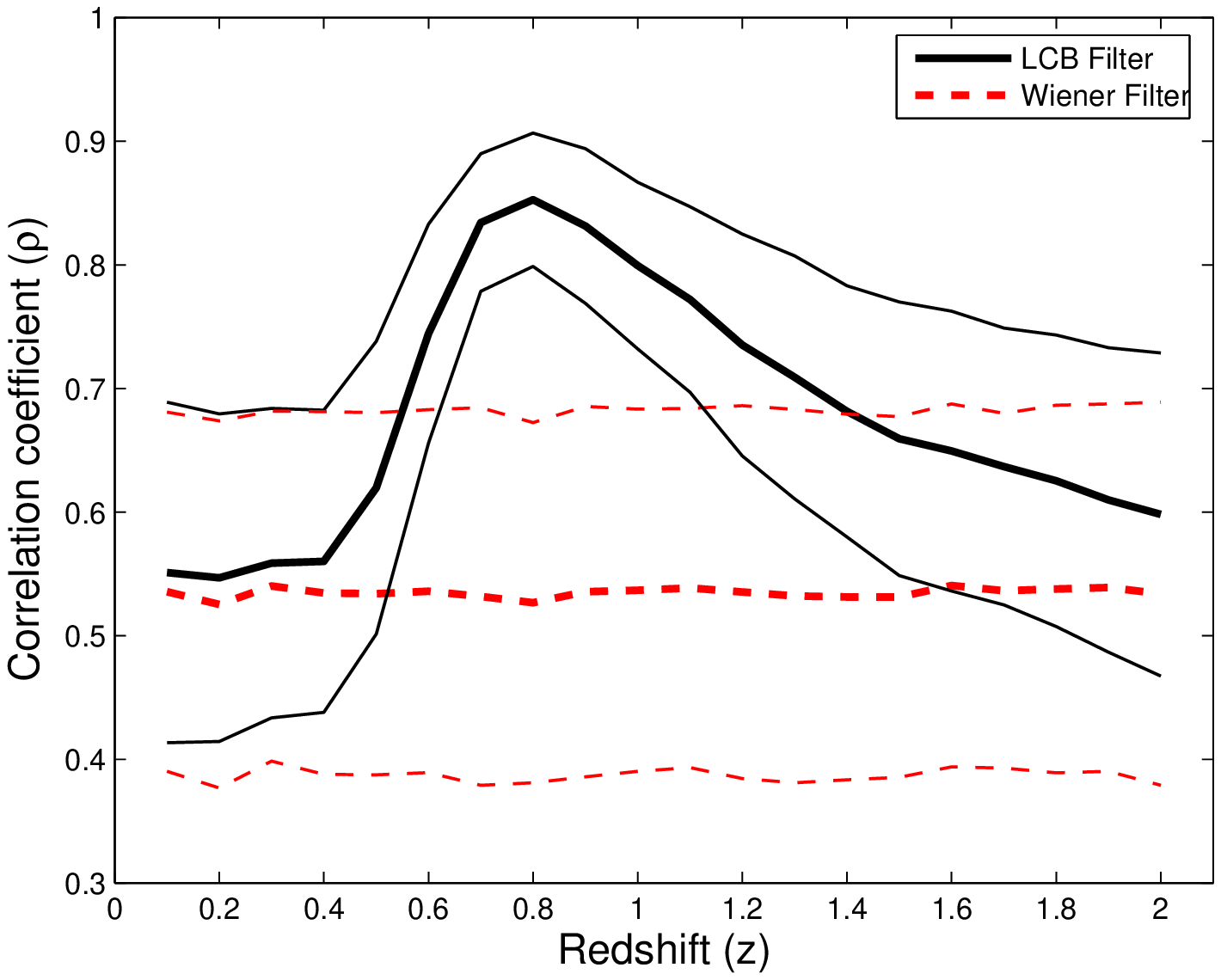}
\caption{Mean correlation coefficient for the ISW reconstruction using
the LCB filter obtained from 1000 simulations (black solid thick
line) and the corresponding 1$\sigma$ region (black thin lines) for
different galaxy surveys (considering the case of the wider bin in
$z$). The left panel corresponds to reconstructions obtained from data
where a mask is applied (discarding around one-fourth of the sky) and
the right panel shows the results obtained assuming a {\emph wrong}
cosmological model. For comparison, the mean correlation coefficient
and the corresponding 1$\sigma$ region for the WF reconstruction are also shown (red dashed lines).}
\label{fig:errors_robust}
\end{figure*}

The top panel of Fig.~\ref{fig:rec_robust} shows the reconstruction of
the ISW map using the LCB filter from the simulated data of
Fig.~\ref{fig:maps} after applying the considered mask. The
cross-correlation coefficient $\rho$, obtained outside the masked
region, is 0.81 to be compared to $\rho$=0.88 in the ideal case. A
more detailed study of the effect of the mask on the quality of the
reconstruction is given in the left panel of
Fig.~\ref{fig:errors_robust}, where the mean and dispersion of the
correlation coefficient for the ISW reconstruction is shown for the
LCB filter (black solid lines) and for the WF (red dashed lines),
considering the case of the wider bin in $z$. This should be compared
to the results obtained for the same case assuming ideal conditions,
given in the left panel of Fig.~\ref{fig:errors}. We can see that the
presence of the mask reduces, although only at the level of a few per
cent, the correlation between the input and reconstructed maps. For
instance, for the best reconstruction, corresponding to a catalogue of
galaxies centred at $z=0.8$, the mean value of the correlation
coefficient is reduced from 0.85, in the ideal case, to 0.81, when the
mask is present. Therefore, the performance of the LCB filter is
robust against incomplete sky observations. Regarding WF,
the mean correlation coefficient is very similar to the one obtained
in the ideal case.

Another interesting issue is the robustness of the method against
uncertainties in the knowledge of the covariance matrix $C(\ell)$ or,
for the considered case of the ISW reconstruction, in the knowledge of
the cosmological model. In the previous examples, we have used the
same cosmological model to generate the simulations and to reconstruct
the ISW map. However, in a more realistic case, the assumed
cosmological parameters may deviate, at a certain level, from their
true underlying values.

Different cosmological data sets (such as CMB, LSS galaxy surveys,
primordial Big-Bang nucleosynthesis, measurements of the Hubble
constant or supernovae data) are currently placing strong constraints
on the cosmological parameters, giving rise to a consistent picture of
the universe, the so-called concordance model. Therefore, given the
current constraints, large deviations from the present estimation of
the cosmological parameters are not expected. In order to test the
sensitivity of the LCB filter to possible errors in the assumed
cosmological model, we have produced reconstructions of the ISW map
using a model different from the one present in the simulations (which
was given by the $\Lambda$-CDM model that best fitted the 3-year WMAP
data), i.e, we obtain the reconstructions using a {\emph wrong}
covariance matrix. This second model corresponds to the parameters
given by the $\Lambda$-CDM model that best fits the 5-year WMAP data,
baryon acoustic oscillations in the distribution of galaxies and Type
I supernovae data (see \cite{kom08} and references therein). The
differences between these two models correspond to the level of
uncertainties that one would expect in our present knowledge of the
cosmological parameters. In particular, the value of the dark energy
density $\Omega_\Lambda$, one of the parameters more closely related
to the ISW signal, differs in a $\sim 5$ per cent between both models.

The bottom panel of Fig.~\ref{fig:rec_robust} shows the reconstruction
of the ISW map using the LCB filter, assuming the alternative
cosmological model, from the simulated data of
Fig.~\ref{fig:maps}. Although, in this case, the amplitude of the
reconstructed ISW is a bit larger (due to the different assumed
cosmological model), the correlation coefficient ($\rho=0.87$) is
still very similar to the value found in the ideal case. In addition,
the right panel of Fig.~\ref{fig:errors_robust} shows the mean and
dispersion of the correlation coefficient for the ISW reconstruction
in this case, averaged over a large number of simulations and
considering the case of the wider bin in $z$. Again, the figure should
be compared to the same case assuming ideal conditions, given in the
left panel of Fig.~\ref{fig:errors}. From the comparison, it is
apparent that the assumption of a different cosmological model does not
affect significantly the mean correlation coefficient between input
and reconstructed ISW map neither for the LCB filter nor for the
WF. This result shows that our methodology is also robust
against the possible uncertainties in our knowledge of the
cosmological parameters.

\section{Conclusions}
\label{sec:conclusions}

We have presented a linear filter, the LCB filter, which reconstructs
a signal from a confusion background taking into account additional
information from a second observation correlated with the sought
signal. To study the performance of the technique, we have applied it
to simulated CMB and LSS observations with the aim of recovering the
ISW effect. This is a very challenging problem, due to the weakness of
the ISW signal and, as far as we know, this is the first time that a
technique is presented that attempts to recover an ISW map instead of
searching for a statistical detection of the signal. We have
considered different possibilities for the LSS survey, in order to
illustrate the dependence of the method on the characteristics of the
chosen catalogue of galaxies. From the considered cases, the best
reconstruction (mean $\rho = 0.85$) is obtained for a survey with a
wide bin in $z$ and centred in $z=0.8$. For comparison, we have also
recovered the ISW signal from CMB simulated maps using a classical WF,
finding that the LCB filter provides better reconstructions. This
confirms the idea that combining information from the CMB and the LSS
survey is useful to recover the ISW map. We have also shown that our
methodology is robust when dealing with incomplete sky observations
and against the possible uncertainties in our knowledge of the
cosmological model. In a future work we will present a more detailed
application to the ISW effect that takes into account more realistic
physical conditions for the galaxy populations. Finally, we would like
to point out that the formalism can be easily generalised to include
information from several galaxy surveys at the same time.

\section*{Acknowledgement}
RBB, PV and EMG acknowledge partial financial support from the Spanish
MEC project AYA2007-68058-C03-02. CHM acknowledges the hospitality of
IFCA at Santander (Spain), where part of this work was carried
out. Some of the results in this paper have been derived using the
HEALPix \cite{gor05} package.

\ifCLASSOPTIONcaptionsoff
  \newpage
\fi

\bibliographystyle{IEEEtran}
\bibliography{IEEEabrv,refs}

\begin{thebibliography}{10}
\providecommand{\url}[1]{#1}
\csname url@samestyle\endcsname
\providecommand{\newblock}{\relax}
\providecommand{\bibinfo}[2]{#2}
\providecommand{\BIBentrySTDinterwordspacing}{\spaceskip=0pt\relax}
\providecommand{\BIBentryALTinterwordstretchfactor}{4}
\providecommand{\BIBentryALTinterwordspacing}{\spaceskip=\fontdimen2\font plus
\BIBentryALTinterwordstretchfactor\fontdimen3\font minus
  \fontdimen4\font\relax}
\providecommand{\BIBforeignlanguage}[2]{{%
\expandafter\ifx\csname l@#1\endcsname\relax
\typeout{** WARNING: IEEEtran.bst: No hyphenation pattern has been}%
\typeout{** loaded for the language `#1'. Using the pattern for}%
\typeout{** the default language instead.}%
\else
\language=\csname l@#1\endcsname
\fi
#2}}
\providecommand{\BIBdecl}{\relax}
\BIBdecl

\bibitem{cha05}
A.~{Challinor}, ``{Cosmic Microwave Background Anisotropies},'' in \emph{The
  Physics of the Early Universe}, ser. Lecture Notes in Physics, Berlin
  Springer Verlag, K.~{Tamvakis}, Ed., vol. 653, 2005, pp. 71--+.

\bibitem{mar06}
E.~{Mart{\'{\i}}nez-Gonz{\'a}lez} and P.~{Vielva}, ``{The Cosmic Microwave
  Background Anisotropies: Open Problems},'' in \emph{The Many Scales in the
  Universe: JENAM 2004 Astrophysics Reviews}, J.~C. {Del Toro Iniesta}, E.~J.
  {Alfaro}, J.~G. {Gorgas}, E.~{Salvador-Sole}, and H.~{Butcher}, Eds., Jan.
  2006, pp. 1--4020.

\bibitem{sac67}
R.~K. {Sachs} and A.~M. {Wolfe}, ``{Perturbations of a Cosmological Model and
  Angular Variations of the Microwave Background},'' \emph{\apj}, vol. 147, pp.
  73--+, Jan. 1967.

\bibitem{pee03}
P.~J. {Peebles} and B.~{Ratra}, ``{The cosmological constant and dark
  energy},'' \emph{Reviews of Modern Physics}, vol.~75, pp. 559--606, Apr.
  2003.

\bibitem{cri96}
R.~G. {Crittenden} and N.~{Turok}, ``{Looking for a Cosmological Constant with
  the Rees-Sciama Effect},'' \emph{Physical Review Letters}, vol.~76, pp.
  575--578, Jan. 1996.

\bibitem{fos03}
P.~{Fosalba}, E.~{Gazta{\~n}aga}, and F.~J. {Castander}, ``{Detection of the
  Integrated Sachs-Wolfe and Sunyaev-Zeldovich Effects from the Cosmic
  Microwave Background-Galaxy Correlation},'' \emph{\apjl}, vol. 597, pp.
  L89--L92, Nov. 2003.

\bibitem{bou04}
S.~{Boughn} and R.~{Crittenden}, ``{A correlation between the cosmic microwave
  background and large-scale structure in the Universe},'' \emph{\nat}, vol.
  427, pp. 45--47, Jan. 2004.

\bibitem{fos04}
P.~{Fosalba} and E.~{Gazta{\~n}aga}, ``{Measurement of the gravitational
  potential evolution from the cross-correlation between WMAP and the APM
  Galaxy Survey},'' \emph{\mnras}, vol. 350, pp. L37--L41, May 2004.

\bibitem{afs04}
N.~{Afshordi}, Y.-S. {Loh}, and M.~A. {Strauss}, ``{Cross-correlation of the
  cosmic microwave background with the 2MASS galaxy survey: Signatures of dark
  energy, hot gas, and point sources},'' \emph{\prd}, vol.~69, no.~8, pp.
  083\,524--+, Apr. 2004.

\bibitem{nol04}
M.~R. {Nolta}, E.~L. {Wright}, L.~{Page}, C.~L. {Bennett}, M.~{Halpern},
  G.~{Hinshaw}, N.~{Jarosik}, A.~{Kogut}, M.~{Limon}, S.~S. {Meyer}, D.~N.
  {Spergel}, G.~S. {Tucker}, and E.~{Wollack}, ``{First Year Wilkinson
  Microwave Anisotropy Probe Observations: Dark Energy Induced Correlation with
  Radio Sources},'' \emph{\apj}, vol. 608, pp. 10--15, Jun. 2004.

\bibitem{vie06}
P.~{Vielva}, E.~{Mart{\'{\i}}nez-Gonz{\'a}lez}, and M.~{Tucci},
  ``{Cross-correlation of the cosmic microwave background and radio galaxies in
  real, harmonic and wavelet spaces: detection of the integrated Sachs-Wolfe
  effect and dark energy constraints},'' \emph{\mnras}, vol. 365, pp. 891--901,
  Jan. 2006.

\bibitem{pie06}
D.~{Pietrobon}, A.~{Balbi}, and D.~{Marinucci}, ``{Integrated Sachs-Wolfe
  effect from the cross correlation of WMAP 3year and the NRAO VLA sky survey
  data: New results and constraints on dark energy},'' \emph{\prd}, vol.~74,
  no.~4, pp. 043\,524--+, Aug. 2006.

\bibitem{mce07}
J.~D. {McEwen}, P.~{Vielva}, M.~P. {Hobson}, E.~{Mart{\'{\i}}nez-Gonz{\'a}lez},
  and A.~N. {Lasenby}, ``{Detection of the integrated Sachs-Wolfe effect and
  corresponding dark energy constraints made with directional spherical
  wavelets},'' \emph{\mnras}, vol. 376, pp. 1211--1226, Apr. 2007.

\bibitem{mce08}
J.~D. {McEwen}, Y.~{Wiaux}, M.~P. {Hobson}, P.~{Vandergheynst}, and A.~N.
  {Lasenby}, ``{Probing dark energy with steerable wavelets through correlation
  of WMAP and NVSS local morphological measures},'' \emph{\mnras}, vol. 384,
  pp. 1289--1300, Mar. 2008.

\bibitem{rac08}
A.~{Raccanelli}, A.~{Bonaldi}, M.~{Negrello}, S.~{Matarrese}, G.~{Tormen}, and
  G.~{de Zotti}, ``{A reassessment of the evidence of the Integrated
  Sachs-Wolfe effect through the WMAP-NVSS correlation},'' \emph{\mnras}, vol.
  386, pp. 2161--2166, Jun. 2008.

\bibitem{gia08}
T.~{Giannantonio}, R.~{Scranton}, R.~G. {Crittenden}, R.~C. {Nichol}, S.~P.
  {Boughn}, A.~D. {Myers}, and G.~T. {Richards}, ``{Combined analysis of the
  integrated Sachs-Wolfe effect and cosmological implications},'' \emph{ArXiv
  e-prints}, vol. 801, Jan. 2008.

\bibitem{wie49}
D.~{Wiener}, \emph{{Extrapolation and Smoothing of Stationary Time
  Series}}.\hskip 1em plus 0.5em minus 0.4em\relax New York, Wiley, 1949.

\bibitem{ski89}
J.~{Skilling}, Ed., \emph{{Maximum entropy and bayesian methods : 8 : 1988}},
  1989.

\bibitem{del08}
J.~{Delabrouille} and J.-F. {Cardoso}, ``{Diffuse source separation in CMB
  observations},'' \emph{ArXiv Astrophysics e-prints}, Feb. 2007.

\bibitem{bou98}
S.~P. {Boughn}, R.~G. {Crittenden}, and N.~G. {Turok}, ``{Correlations between
  the cosmic X-ray and microwave backgrounds: constraints on a cosmological
  constant},'' \emph{New Astronomy}, vol.~3, pp. 275--291, Jul. 1998.

\bibitem{sel96}
U.~{Seljak} and M.~{Zaldarriaga}, ``{A Line-of-Sight Integration Approach to
  Cosmic Microwave Background Anisotropies},'' \emph{\apj}, vol. 469, pp.
  437--+, Oct. 1996.

\bibitem{spe07}
D.~N. {Spergel}, R.~{Bean}, O.~{Dor{\'e}}, M.~R. {Nolta}, C.~L. {Bennett},
  J.~{Dunkley}, G.~{Hinshaw}, N.~{Jarosik}, E.~{Komatsu}, L.~{Page}, H.~V.
  {Peiris}, L.~{Verde}, M.~{Halpern}, R.~S. {Hill}, A.~{Kogut}, M.~{Limon},
  S.~S. {Meyer}, N.~{Odegard}, G.~S. {Tucker}, J.~L. {Weiland}, E.~{Wollack},
  and E.~L. {Wright}, ``{Three-Year Wilkinson Microwave Anisotropy Probe (WMAP)
  Observations: Implications for Cosmology},'' \emph{\apjs}, vol. 170, pp.
  377--408, Jun. 2007.

\bibitem{ben03}
C.~L. {Bennett}, R.~S. {Hill}, G.~{Hinshaw}, M.~R. {Nolta}, N.~{Odegard},
  L.~{Page}, D.~N. {Spergel}, J.~L. {Weiland}, E.~L. {Wright}, M.~{Halpern},
  N.~{Jarosik}, A.~{Kogut}, M.~{Limon}, S.~S. {Meyer}, G.~S. {Tucker}, and
  E.~{Wollack}, ``{First-Year Wilkinson Microwave Anisotropy Probe (WMAP)
  Observations: Foreground Emission},'' \emph{\apjs}, vol. 148, pp. 97--117,
  Sep. 2003.

\bibitem{con98}
J.~J. {Condon}, W.~D. {Cotton}, E.~W. {Greisen}, Q.~F. {Yin}, R.~A. {Perley},
  G.~B. {Taylor}, and J.~J. {Broderick}, ``{The NRAO VLA Sky Survey},''
  \emph{\aj}, vol. 115, pp. 1693--1716, May 1998.

\bibitem{kom08}
E.~{Komatsu}, J.~{Dunkley}, M.~R. {Nolta}, C.~L. {Bennett}, B.~{Gold},
  G.~{Hinshaw}, N.~{Jarosik}, D.~{Larson}, M.~{Limon}, L.~{Page}, D.~N.
  {Spergel}, M.~{Halpern}, R.~S. {Hill}, A.~{Kogut}, S.~S. {Meyer}, G.~S.
  {Tucker}, J.~L. {Weiland}, E.~{Wollack}, and E.~L. {Wright}, ``{Five-Year
  Wilkinson Microwave Anisotropy Probe (WMAP) Observations: Cosmological
  Interpretation},'' \emph{ArXiv e-prints}, vol. 803, Mar. 2008.

\bibitem{gor05}
K.~M. {G{\'o}rski}, E.~{Hivon}, A.~J. {Banday}, B.~D. {Wandelt}, F.~K.
  {Hansen}, M.~{Reinecke}, and M.~{Bartelmann}, ``{HEALPix: A Framework for
  High-Resolution Discretization and Fast Analysis of Data Distributed on the
  Sphere},'' \emph{\apj}, vol. 622, pp. 759--771, Apr. 2005.

\end{thebibliography}

%
\begin{IEEEbiography}
{R.B. Barreiro}
obtained her B.S. degree in 1995 from the Universidad
de Santiago de Compostela, Spain, completing also as part of her
degree one year at the University of Sheffield, UK. She completed her
Ph.D. in astrophysics in the Universidad de Cantabria in 1999. After
her Ph.D., she worked as a Research Associate at the Cavendish
Laboratory of the University of Cambridge, UK. In 2002, she
moved to the Instituto de F{\'\i}sica de Cantabria (IFCA, CSIC-UC),
Spain, with a long term contract. Currently, she continues at IFCA as
a tenured scientist of the Universidad de Cantabria. Her research
interests are mainly focused in the field of the cosmic microwave
background, including statistical data analysis, in particular the
study of the Gaussianity of the CMB, and the development of component
separation techniques for both diffuse emissions and compact sources.
\end{IEEEbiography}

\begin{IEEEbiography}
{P. Vielva}
is a \emph{Ram\'on y Cajal} researcher of the Instituto de
F{\'\i}sica de Cantabria (IFCA, CSIC-UC). He obtained his B.S.
degree in 1998 from the Universidad de Cantabria. He completed his
Ph.D., at IFCA. In 2004 he was a postdoctoral researcher at the
Coll{\`e}ge de France/APC in Paris. In 2005 he moved to IFCA were he
was a researcher of the Universidad de Cantabria associated to the
ESA Planck mission. In 2006 he obtained an CSIC-I3P position at IFCA
and during that year he was a visiting researcher at the Cavendish
Laboratory of the University of Cambridge. His researching field is
cosmology, specially the study of the cosmic microwave background
(CMB). In particular, he works on the development and application of
statistical and signal processing tools for the CMB data analysis.
\end{IEEEbiography}

\begin{IEEEbiography}
{C. Hern\'andez-Monteagudo}
is currently a post-doc at the Max Planck Institut f\"ur Astrophysik
in Munich (Germany). He finished his Ph.D. in Physics at the
University of Salamanca (Spain) in February 2002, and since then he
has been doing research at the Max Planck Institut f\"ur Astrophysik
in Munich and the University of Pennsylvania (USA). His interests span
different aspects of the study of the Cosmic Microwave Background
(CMB) radiation, such as cosmological recombination, reionization,
pollution of the Inter Galactic Medium with the first metals,
Comptonization of the CMB spectrum via the thermal Sunyaev-Zel'dovich
effect, and other secondary effects such as kinetic Sunyaev-Zel'dovich
effect, Rees Sciama and Integrated Sachs-Wolfe effect.  Currently he
is involved in three major projects in Cosmology: Planck (an ESA
mission that attempts to measure CMB intensity and polarization
anisotropies with unprecedented sensitivity and systematic control),
the Atacama Cosmology Telescope (a collaboration in the US probing the
small scales of the CMB angular power spectrum) and the PAU-BAU
consolider project (a Spanish collaboration devoted to the study of
the nature of Dark Energy).
\end{IEEEbiography}

\begin{IEEEbiography}
{E. Mart\'\i nez-Gonz\'alez} 
is currently a Research Professor at the Instituto de F\'\i sica de
Cantabria (IFCA, CSIC-UC). He obtained his B.S. degree in the
Universidad de Cantabria in 1986 and followed his doctorate studies at
NORDITA (Copenhagen, Denmark). He was the coordinator of the first
European Research Network on the CMB funded by the EU. Currently, he
is Co-Investigator of the Planck Low Frequency Instrument consortium.
His research fields cover different theoretical and observational
aspects of Cosmology, mainly focusing on the Cosmic Microwave
Background.
\end{IEEEbiography}




\end{document}